\pdfoutput=1
\def\year{2020}\relax
\documentclass[letterpaper]{article} % DO NOT CHANGE THIS
\usepackage{aaai20}  % DO NOT CHANGE THIS
\usepackage{times}  % DO NOT CHANGE THIS
\usepackage{helvet} % DO NOT CHANGE THIS
\usepackage{courier}  % DO NOT CHANGE THIS
\usepackage[hyphens]{url}  % DO NOT CHANGE THIS
\usepackage{graphicx} % DO NOT CHANGE THIS
\usepackage{graphicx}  % DO NOT CHANGE THIS
\usepackage[normalem]{ulem}

\title{Computer-Generated Music for Tabletop Role-Playing Games}
\author {
	Lucas N. Ferreira,\textsuperscript{\rm 1} Levi H. S. Lelis,\textsuperscript{\rm 2} Jim Whitehead\textsuperscript{\rm 1} \\
	\textsuperscript{\rm 1}Department of Computational Media, University of California, Santa Cruz, USA \\
	\textsuperscript{\rm 2}Department of Computing Science, Alberta Machine Intelligence Institute (Amii), University of Alberta, Canada \\
	lferreira@ucsc.edu, levi.lelis@ualberta.ca, ejw@soe.ucsc.edu \\
	}

\newcommand{\cor}[1]{
\cellcolor{orange!40}
}

\usepackage{amsfonts}
\usepackage{amsmath}
\usepackage[noend]{algorithmic}
\usepackage{algorithm}
\usepackage{listings}
\usepackage[table]{xcolor}
\usepackage{caption}
\usepackage{subcaption}
\usepackage{multirow}
\usepackage{graphicx,subcaption}
\usepackage{stmaryrd}
\usepackage{booktabs}
\usepackage[textsize=tiny,textwidth=0.55in]{todonotes}
\begin{document}

\maketitle

\begin{abstract}
In this paper we present Bardo Composer, a system to generate background music for tabletop role-playing games. Bardo Composer uses a speech recognition system to translate player speech into text, which is classified according to a model of emotion. Bardo Composer then uses Stochastic Bi-Objective Beam Search, a variant of Stochastic Beam Search that we introduce in this paper, with a neural model to generate musical pieces conveying the desired emotion.
%The system uses a Transformer neural network to classify the emotion of the game story based on the players' speeches. Another Transformer is used to classify the emotion of MIDI piano pieces. Both these classifiers are used to control the emotion of pieces generated by a third neural model. We evaluate the story classifier using the Call of the Wild dataset and results show that it outperforms previous approaches for classifying the emotion of the players' speeches emotions by approximately 10\% in terms of accuracy.
We performed a user study with 116 participants to evaluate whether people are able to correctly identify the emotion conveyed in the pieces generated by the system. In our study we used pieces generated for Call of the Wild, a Dungeons and Dragons campaign available on YouTube. Our results show that human subjects could correctly identify the emotion of the generated music pieces as accurately as they were able to identify the emotion of pieces written by humans.
%how much the generated pieces match the emotion in the story.\todo{Add summary of results.}
\end{abstract}

\section{Introduction}

In this paper we introduce Bardo Composer, or Composer for short, a system for generating
musical pieces that match the emotion of
stories told in tabletop role-playing games (TRPGs). For example, if the players are fighting a
dragon, Composer should generate a piece matching such an epic moment of the story. TRPG players often manually choose songs to play as background music to enhance their experience~\cite{bergstrom2014case}. Our goal is to develop an intelligent system that augments the players' experience with soundtracks that match the story being told in the game.
Importantly, the system should allow players to concentrate on the role-playing part of the game, and not on the disruptive task of selecting the next music piece to be played.
The object of our research is Dungeons and Dragons (D\&D), a TRPG where players interpret characters of a story conducted by a special player called the dungeon master.

\citeauthor{padovani2017}~\shortcite{padovani2017,padovaniFL19} introduced Bardo, a system that automatically selects the background music of a D\&D session based on the story being told by the players. This paper builds upon their system. Bardo uses a speech recognition system to transcribe voice into text, which is then classified into an emotion. Bardo then selects a song of the classified emotion from a library of labeled songs. In this work we extend Bardo to include a neural model for generating musical pieces conveying the emotions detected in the game story, instead of selecting a song from a labeled library---thus the name Bardo Composer.
%The motivation for generating pieces
%as opposed to selecting pieces
%is that they
We expect that by generating pieces we can capture the exact emotional tone of the story, while methods that select from a set of pre-composed pieces have a more limited ``emotional palette''.

% When one can only select from a limited amount of pre-compose pieces, the ``emotional palette'' might be limited as compared to a system that can generate a wide range of pieces. % in real time.

%Previous works have investigated the problem of music generation and
Language models (LMs) are able to generate coherent music pieces~\cite{ferreira_2019}. However, it is still
challenging
%to control such models
to generate music with a given emotion.
%in addition to being of good quality.
For that we introduce Stochastic Bi-Objective Beam Search (\texttt{SBBS}), a variant of Stochastic Beam Search \cite{poole2010artificial} to guide the generative process while maximizing the probability given by a LM jointly with the probability of pieces matching an emotion. % in the story.
%We use a a fine-tuned BERT \cite{devlin2018bert} for classifying the emotion in the story.
The emotion in the story is detected by a
%fine-tuned
BERT model \cite{devlin2018bert} and is given
as input to \texttt{SBBS}, which uses a GPT-2 model \cite{radford2019language}  to classify the emotion of
the generated pieces.

We evaluated Composer on the Call of the Wild (CotW) dataset~\cite{padovani2017}, which is a campaign of D\&D available on YouTube. Since our primary goal is to generate music pieces conveying the current emotion of the game's story, we used Composer to generate pieces of parts of CotW that featured a transition in the story's emotion. Then, in a user study with 116 participants, we evaluated whether people correctly perceive the intended emotions in pieces generated by Composer.
%We used Bardo's original scheme as a baseline in our study.
%That is,
We also measured if the participants were able to distinguish the emotion of human-composed pieces by using Bardo's original system as a baseline. Our results show that the participants were able to identify the emotions in generated pieces as accurately as they were able to identify emotions in human-composed pieces. This is an important result towards the goal of a fully-automated music composition system for TRPGs.

\section{Related Work}

Our work is mostly related to machine learning models that generate music with a given emotion. For example,
\citeauthor{monteith2010automatic}~\shortcite{monteith2010automatic}
trained Hidden Markov models to generate music from a corpus labeled
according to a categorical model of emotion. These models are trained for
each emotion to generate melodies and underlying harmonies.
%with rhythms
%being sampled randomly from examples of a given emotion.
Ferreira and Whitehead \shortcite{ferreira_2019} used a genetic algorithm to fine-tune a pre-trained LSTM,
%The LR model highlights the subset of neurons that carry sentiment signal, so a genetic algorithm was used to fine tune those neurons,
controlling the LSTM to generate either positive or negative pieces.
Our work differs from \citeauthor{monteith2010automatic}'s because we train a single LM
that is controlled to generate music with different emotions. It is also different
from \cite{ferreira_2019} once we control
the LM at sampling time and not at training
time.

% This paper is inspired by the system Bardo \cite{}, which selects music
% for tabletop games.
% \todo{Add related work from ISMIR paper. What is currently in this section isn't probably necessary because we talk about this work in various parts of the paper.}
%This paper is related to systems designed to generate music with a given emotion.
Our work is also related to rule-based systems that map musical features
to a given emotion \cite{williams2015investigating}.
For example, \citeauthor{williams2015dynamic}~\shortcite{williams2015dynamic} generate
soundtracks for video games using a rule-based system to transform pre-generated melodies,
matching the emotion of annotated game scenes. % annotated according to a model of emotion.
\citeauthor{davis2014generating}~\shortcite{davis2014generating}
follow a similar approach in TransPose, a system that generates piano melodies
for novels. %TransPose uses a lexicon-based approach to automatically detect categorical emotions in novels and a rule-based technique to create piano melodies with
%these emotions.
Our work differs from these rule-based systems because we learn mappings from musical features to emotion directly from data. %, instead of hard-coded rules.
% More specifically, the wcork of Ferreira
% and Whitehead \cite{} is very related to this paper, once
% we extend their approach to generate music with emotion. Ferreira
% and Whitehead pre-trained a GPT-2 transformer as a LM
% on the piano tracks of the Lakh Midi dataset \cite{} and fine-tuned it
% as a classifier of sentiment in music (positive or negative) using
% the VGMIDI dataset \cite{}. Ferreira and Whitehead used this fine-tuned
% classifier to steer the distribution given by LM towards
% a desired sentiment.
%Due to the lack of labeled data,
%There is a limited set of machine
%learning models to generate music with emotion. For example,

%In a general way, this paper
Our work is also related
to neural models that generate text with a given characteristic.
% using neural networks,
% which we think can be generalized to music and thus are even more
% relevant to this paper.
For example, CTRL \cite{keskar2019ctrl} is a
Transformer LM trained to generate text conditioned on
special tokens that inform the LM about the characteristics of the text to be generated (e.g., style).
% These control codes are derived automatically from the origin of
% the text (e.g., Wikipedia). %, which means no manual annotation had to be performed.
% During training, every example is fed as input together with
% its control codes.
%CTRL can be used to generate text with a
%particular style, for example. % by feeding the information that represents that style.
Our work differs
from CTRL because we control the LM with
a search procedure and not with an extra input to the LM. Conditioning the LM requires a large amount of labeled data, which is expensive
in our domain. % the music and emotion domain.

% \citeauthor{ziegler2019fine}~\shortcite{ziegler2019fine}
% used reinforcement
% learning (RL) to fine-tune a Transformer LM for generating text with a given sentiment by using a reward model trained
% on human preferences on text continuations. %\citeauthor{ziegler2019fine} fine-tuned the model
% %for two different purposes: generating text with sentiment and text
% %summarization.
% Our work differs from \citeauthor{ziegler2019fine}'s because we control the LM at sampling time (with search) and not at training time (with RL).
% %They achieve good results with only 5,000 comparisons evaluated by humans.

The Plug and Play LM
\cite{dathathri2019plug} combines a pre-trained LM
with a small attribute classifier to guide text generation.
%without any  further training of the LM.
%This is achieved by shifting
%the hidden layers of the pre-trained LM at each time-step of
%the generation process.
Although both Composer and the Plug and Play LM control the generation procedure at sampling time, we use search as a means of generation control while Plug and Play LM uses a classifier to alter the structure of the model. %updates the hidden neurons.
% This shift updates the hidden layers
% towards the direction of the sum of two gradients: one towards higher
% log-likelihood of the attribute and one towards higher log-likelihood of the LM.

%Our work is also related to Stochastic Beam Search for sequence generation. For example,

\citeauthor{vijayakumar2018diverse}~\shortcite{vijayakumar2018diverse} and \citeauthor{Kool2019SBS}~\shortcite{Kool2019SBS} proposed variations of Beam search
%(DBS) to solve the problem that generated sentences are often minor re-wordings of a common utterance.
%DBS decodes diverse lists by dividing candidate solutions into groups and enforcing diversity between groups.
%\citeauthor{Kool2019SBS}~\shortcite{Kool2019SBS} proposed a Beam search algorithm
to solve the problem of generating repetitive sentences.
%by applying
%the ``Gumbel-Top-k'' trick to sample without replacement with Beam Search, mitigating the problem of generating repetitive sequences.
Our work differs from both these works because our variation of Beam search optimizes for two independent objectives.

\section{Background}

%\subsection{Symbolic Music Composition}
\vspace{0.05in}
\noindent
\textbf{Symbolic Music Composition} Symbolic music is typically generated by sampling from a LM that computes the likelihood of the next musical symbols (e.g., note) in a piece. Typically,
the LM is defined as a neural network and the symbols are extracted from MIDI or piano roll representations of music ~\cite{briot2017deep}. Let $x = [x_0, \cdots, x_{t-2}, x_{t-1}]$ be the first $t$ symbols of a piece and $P(x_t|{x_0, \cdots, x_{t-2}, x_{t-1}})$ be the probability of the next symbol being $x_t$, according to a trained LM. One can sample the next symbol of the sequence according to the probability distribution $P$~\cite{briot2017deep}. We denote the trained language
model as $L$ and $L(x)$ is a function that returns the next symbol given a sequence $x$.
%given a sequence of previous symbols $\{x^{t-1}, x^{t-2}, ..., x^0\}$
%(i.e. LMs) \cite{}.
To generate a piece with $L$, one %can either provide an empty sequence $x$ as input or
provides as input a sequence of symbols $x = [x_0, x_1, \cdots, x_t]$ to bias the generation process.
%starts with an input
%sequence .
This input sequence
is fed into $L$ which computes $L(x) = x_{t + 1}$.
Next, $x_{t + 1}$ is concatenated with $x$ and the process
repeats until a special end-of-piece symbol is found or
a given number of symbols are generated.

% In this paper %we assume music pieces from a MIDI dataset
% %encoded using the scheme proposed by Ferreira and Whitehead \cite{}.
% we use a MIDI encoding
% %In this encoding scheme, a MIDI file is represented as
% %sequence $x = [x_0, x_1, ..., x_n]$,
% where each $x_i$ in a sequence $x$ is a
% symbol from a vocabulary~\cite{ferreira_2019}, which we will denote as $V$. The 149 symbols present in $V$ are divided into four types.
% \begin{enumerate}
%     \item  $n_i$: play the note $n$ with pitch $30 < i < 96$.
%     \item $d_i$ set the duration in timesteps of the notes to $0 < i < 56$
%     \item $v_i$: set the velocity (in MIDI units) of the remaining notes in the sequence.
%     \item TS and END: determines the end-of-timestep and end-o-piece, respectively.
% \end{enumerate}

%\subsection{Bardo}

\vspace{0.05in}
\noindent
\textbf{Bardo} \citeauthor{padovani2017}~\shortcite{padovani2017,padovaniFL19} presented Bardo, a system to select background music for tabletop games. Bardo  classifies sentences produced by a speech recognition system into one of the
four story emotions: Happy, Calm, Agitated, and Suspenseful. Bardo then selects a song from a
library of songs corresponding to the classified emotion.
The selected song is then played as background
%music whenever NB detects an emotion transition in the story.
music at the game table. %\jarvis\ operates in real time and
In this paper we use Padovani et al.'s dataset to train an emotion classifier for the story being told at a game session. Their dataset includes 9 episodes of CotW, which contains 5,892 sentences and 45,247 words, resulting in 4 hours, 39 minutes, and 24 seconds of gameplay. There are 2,005 Agitated, 2,493 Suspenseful, 38 Happy, and 1,356 Calm sentences in the dataset.

%\subsection{Valence-Arousal Model of Emotion}

%To generate a piece for a sentence, we first classify the emotion of that
%sentence using a Transformer neural network~\cite{}.
\vspace{0.05in}
\noindent
\textbf{Valence-Arousal Model of Emotion} We use a two-dimensional emotion model that generalizes the emotion model used in Bardo.
%The emotion of the sentence
%is classified according to a discrete
We consider the dimensions of valence and arousal, denoted by a pair $(v,a)$, where $v \in [0, 1]$
and $a \in [0, 1]$~\cite{russell1980circumplex}.
% This is a two-dimensional model that represents
% an emotion with valence-arousal pair $(v,a)$, where $v \in [0, 1]$
% and $a \in [0, 1]$.
Valence measures sentiment and thus $v = 0$ means a negative input and $v = 1$ means a positive input. Arousal measures the
energy of the input and thus $a = 0$ means that the input has low energy whereas $a = 1$ means that the input has high energy. We use this model for classifying both the emotion of the player's speeches and the emotion of the generated music.

\section{Bardo Composer: System Description}

\begin{algorithm}[t]
\caption{Bardo Composer}
\label{alg:bardo}
\begin{algorithmic}[1]
\REQUIRE Speech recognition system $S$, Text emotion classifier $E_s$, Music emotion classifier $E_m$, LM $L$, speech signal $v$, previously composed symbols $x$, beam size $b$, number of symbols $k$
\ENSURE Music piece $x$
\STATE $s, l \gets S(v)$ \label{line:voice2text}
\STATE $v, a \gets E_s(s)$ \label{line:emotion_classification}
% \STATE $x \gets \{\}$ \label{line:init}
%\STATE $i \gets 0$
%\WHILE{time of play of $[x_{t}, \cdots, x_{t + i}]$ is lower than $k$} \label{line:generation1}
\STATE $y \gets$ \texttt{SBBS}$(L, E_m, x, v, a, b, k, l)$ \# \emph{see Algorithm~\ref{alg:sbs}} \label{line:sbs}
\RETURN $x \cup y$ \label{line:generation2}
%\STATE $i \gets i + 1$
%\ENDWHILE
%\RETURN $x$
\end{algorithmic}
\end{algorithm}

A general overview of Composer is shown in Algorithm~\ref{alg:bardo}. It receives as input a speech recognition system $S$, an emotion classifier for text $E_s$, an emotion classifier for music $E_m$, a LM for symbolic music generation $L$, a speech signal $v$ with the last sentences spoken by the players, and a sequence $x$ of musical symbols composed in previous calls to Composer. The algorithm also receives parameters $b$ and $k$, which are used in the search procedure described in Algorithm
~\ref{alg:sbs}. Composer returns a symbolic piece that tries to match the emotion in the players' speeches.

Composer starts by converting the speech signal $v$ into text $s$ with $S$ (line~\ref{line:voice2text}). In addition to text, $S$ returns the duration of the signal $v$ in seconds, this is stored in $l$. Then, Composer classifies the emotion of $s$ in terms of valence $v$ and arousal $a$ and it
%initializes an empty sequence $x$ of musical symbols (line~\ref{line:init}) and it
invokes our Stochastic Bi-Objective Beam Search (\texttt{SBBS}) to generate a sequence of symbols $y$ that matches the desired length $l$ and emotion with arousal $a$ and valence $v$.
%the next symbol in the piece by invoking a Stochastic Beam Search (\texttt{SBBS}) procedure, until the generated piece matches the input voice signal in terms of length (line~\ref{line:generation1}).
\texttt{SBBS} receives as input the models $L$ and $E_m$, the current sequence $x$, the desired emotion values $v$ and $a$, \texttt{SBBS}'s parameter values $b$ and $k$, which are explained below, and the desired length $l$ of the piece to be generated. %\texttt{SBBS} is described below.

In the first call to Composer, the sequence $x$
% can be either empty or contain a pre-composed piece. The idea of providing a pre-composed piece as input is to bias the generation process. In our experiments
is initialized with the the symbols of the first 4 timesteps of a random human-composed piece with the emotion $v, a$, as returned by $E_s$.
% In this case, the excerpt of $x$ containing the human-composed piece
% is not played back to the players, it
Every time
there is a transition from one emotion to another, we reinitialize the sequence $x$ using the same process. This is used to bias the generative process and to emphasize emotion transitions.
%As we explained, an alternative is to initialize $x$ with a sequence of length $t$ of a piece that has the emotions $v$ and $a$ returned by the emotion classifier. This way one can bias the piece generation process (lines~\ref{})

To be used in real-time, Composer is invoked with the most recently captured speech signal $v$ and returns a composed piece of music. While the most recent piece is being played at the game table, Composer receives another signal $v$ and composes the next excerpt. One also needs to define the length of the signal $v$. In our implementation, similar to Padovani et al.~\shortcite{padovani2017}, we use YouTube's subtitle system as the speech recognition system $S$. Therefore, signals $v$ are long enough to form a
%sentence in the form of a
subtitle.

% The main contributions of this paper is the pipeline described in Algorithm~\ref{alg:bardo} and the \texttt{SBBS} search procedure, for generating musical pieces that have a target emotion in terms of valence and arousal. A secondary contribution is an experiment showing that a specially tuned GTP2 classifier can achieve much higher accuracy as classifier $E_s$.

% The system processes sentences spoken by the players as a stream
% (one after the other), generating a piece of music after each sentence
% is processed. Following the approach used in Bardo \cite{}, we extract
% these sentences from the speech of the players using an off-the-shelf
% speech recognition system $S$. Whenever a new sentence $s_i$ is available
% (is spoken), we classify the emotion $e_i$ of this sentence using a text emotion
% classifier $E$. Finally, we use a stochastic beam search method to generate
% a piece of music that matches the emotion $e_i$. This beam search takes as input
% the emotion $e_i$ and the music piece generated in the previous step $x_{i-1}$.
% This process repeats until the game is over. Procedure \ref{alg:system} describes
% this process in pseudo-code.

\subsection{Classifying the Story's Emotion}

In order to have a common model of emotion between stories and music,
we use a mapping from Bardo's four emotion model to the valence-arousal model. Namely, we have Suspenseful mapping to low valence and arousal $(v = 0, a = 0)$; Agitated to low valence and high arousal $(v = 0, a = 1)$; Calm to high valence and low arousal $(v = 1, a = 0)$; and Happy to high valence and arousal $(v = 1, a = 1)$.

% \begin{enumerate}
%     \item Low valence and arousal: $v = 0, a = 0$ (Suspenseful).
%     \item Low valence and high arousal: $v = 0, a = 1$ (Agitated).
%     \item High valence and low arousal: $v = 1, a = 0$ (Calm).
%     \item High valence and arousal: $v = 1, a = 1$ (Happy).
% \end{enumerate}
For example, in the context of the game Dungeons and Dragons, the sentence ``Roll initiative'' is normally said
at the beginning of battles and it can be considered $(v = 0, a = 1)$, once a battle
is a negative (dangerous) moment with high energy. ``Roll initiative'' is normally classified as Agitated in Padovani et al.'s dataset. This mapping allows us to use the valence-arousal model with the labeled CotW dataset.

The valence-arousal mapping is based on the model used to annotate the VGMIDI dataset. When human subjects annotated that dataset, they used a continuous valence/arousal model with labels defining a fixed set of discrete basic emotions (e.g. happy or sad) \cite{ferreira_2019}.

% We use this model
% because music classifiers normally use a similar model, which makes it
% easier to match emotions between the sentences and the music pieces.

Given the limited amount of TRPG stories labeled according
to emotion (there are only 5,892 sentences in the CotW dataset), we use a transfer learning
approach to classify
the sentences~\cite{Radford2018}. We fine-tune a high-capacity BERT architecture with the CotW dataset  \cite{devlin2018bert}. We use BERT because it outperforms
several other transformers across different NLP tasks \cite{devlin2018bert}.
%a specific relatively small dataset of annotated tabletop
%RPG stories.
Although in Algorithm~\ref{alg:bardo} we depict the classifier for story emotions as a single $E_s$ model,
in our implementation we treat valence and arousal independently, thus we
fine-tune a pre-trained BERT for each dimension.

% \subsection{Music Generation with Stochastic Bi-Objective Beam Search}

% Although LMs are able to generate
% music pieces of good quality,
% %such models are unable to generate music with a given emotion.
% it is still challenge to control such models to generate music with a
% given emotion \cite{ferreira_2019}.
% Next, we describe how one can use a LM and a music emotion classifier to bias the process of music generation to match a particular emotion (line~\ref{line:sbs} of Algorithm~\ref{alg:bardo}).

\subsection{Classifying the Music's Emotion}

% We use a music emotion classifier $E_m$ similar to the one introduced by \citeauthor{ferreira_2019}~\shortcite{ferreira_2019} to bias the generation process.

As was the case with the TRPG stories, given the limited amount of MIDI pieces
labeled according to emotion, we also apply a transfer
learning approach to classify emotion in music ($E_m$).
However, different than the $E_s$ model
where we fine-tune a BERT architecture, for $E_m$ we fine-tune a GPT-2 architecture~\cite{radford2019language}.
We use GPT-2 for $E_m$ because it is better suited for sequence generation than BERT.
Similarly to $E_s$, model $E_m$ also treats valence and arousal independently.
Thus, we fine-tune a pre-trained GPT-2 for each of these
dimensions.

% In the text domain, there are several publicly available high-capacity models already pre-trained on very large datasets \cite{devlin2018bert,radford2019language}.
To the best of our knowledge, in the symbolic music domain, there are no publicly available high-capacity LM pre-trained with large (general) datasets. Typically, models in this domain are trained with relatively small and specific datasets. For example, the MAESTRO dataset \cite{hawthorne2018enabling}, the Bach Chorales \cite{hadjeres2017deepbach} and the VGMIDI \cite{ferreira_2019} dataset. We pre-train a general high-capacity GPT-2 architecture as a language model ~\cite{radford2019language} using a new dataset we created called ADL (Augmented Design Lab) Piano MIDI dataset \footnote{https://github.com/lucasnfe/adl-piano-midi}.

The ADL Piano MIDI dataset is based on the Lakh MIDI dataset \cite{raffel2016learning}, which, to the best of our knowledge,
is the largest MIDI dataset publicly available. The Lakh MIDI dataset contains a collection of 45,129 unique MIDI files that
have been matched to entries in the Million Song dataset~\cite{bertin2011million}. Among these files, there are many versions of the same piece. We kept only one version of each piece. Given that the datasets for emotion classification in music are limited to piano only, we extracted from the Lakh MIDI dataset only the tracks with instruments from the ``piano family''(MIDI program numbers 1-8 in the dataset).
This process generated a total of 9,021 unique piano MIDI files. These files are mainly Rock and Classical pieces, so to increase the genre diversity (e.g. Jazz, Blues, and Latin) of the dataset, we
included an additional 2,065 files scraped from public sources on the Internet\footnote{\url{https://bushgrafts.com/midi/} and \url{http://midkar.com/jazz/}}. All files in the final collection were de-duped according to their MD5 checksum. The final dataset has 11,086 pieces.

After pre-training the high-capacity GPT-2 model, we
fine-tune two independent models (one for valence and one for arousal) with an extended version of the VGMIDI dataset~\cite{ferreira_2019}.
% Following the approach of \citeauthor{Radford2018}~\shortcite{Radford2018}, we fine-tune these models by adding and extra layer to the model and training the entire model (including the pre-trained layers)
% with the VGMIDI dataset~\cite{ferreira_2019}.
We extended the VGMIDI dataset from 95 to 200 labeled pieces using the same annotation method of the original dataset.
All the 200 pieces are piano arrangements of video game soundtracks labeled according to the valence-arousal model of emotion.

\subsubsection{Encoding}

We encode a MIDI file by parsing all notes from the
\texttt{NOTE\_ON} and \texttt{NOTE\_OFF}
events in the MIDI. We define a note as a set $z = (z_p, z_s, z_d, z_v)$,
where $\{z_p \in \mathbb{Z} \vert 0 \leq z_p < 128 \}$ is the
pitch number, $\{z_s \in \mathbb{Z} \vert z_s \geq 0 \}$ is the note starting time
in timesteps,  $\{z_d \in \mathbb{Z} \vert 0 \leq z_d \leq 56\}$ is note duration
in timesteps and $\{z_v \in \mathbb{Z} \vert 0 \leq z_v < 128 \}$ is the
note velocity.
Given a MIDI \texttt{NOTE\_ON} event, we parse a note $z$ by retrieving
the starting time $z_s$ (in seconds), the pitch number $z_p$ and
the velocity $z_v$ from that event. To calculate the note duration
$z_d$, we find the correspondent \texttt{NOTE\_OFF} event of the given \texttt{NOTE\_ON}
and retrieve the \texttt{NOTE\_OFF} end time $z_e$ (in seconds). We discretize
$z_s$ and $z_e$ to compute the note duration $z_d = t \cdot z_e - t \cdot z_s$ in timesteps,
where $t$ is a parameter defining the sampling frequency of the timesteps.

We derive a sequence $x = \{z_v^1, z_{d}^1, z_{p}^1, \cdots, z_v^n,
z_{d}^n, z_p^n\}$ of tokens for a given MIDI file by (a)
parsing all notes $z^i$ from the file, (b) sorting them by
starting time $z_s^j$ and (c) concatenating their velocity $z_v^j$,
duration $z_d^j$ and pitch $z_p^j$. We add two special tokens
\texttt{TS} and \texttt{END} in the sequence $x$, to mark
the end of a timestep and the end of a piece, respectively.
This encoding yields a vocabulary $V$ of size $|V| = 314$.
% Figure \ref{fig:encoding} illustrates an example of MIDI file encoded using this approach.

\subsection{Stochastic Bi-Objective Beam Search}

% Although LMs are able to generate
% music pieces of good quality,
% %such models are unable to generate music with a given emotion.
% it is still challenge to control such models to generate music with a
Next, we describe how one can use a LM and a music emotion classifier to bias the process of music generation to match a particular emotion (line~\ref{line:sbs} of Algorithm~\ref{alg:bardo}). For that we introduce Stochastic Bi-Objective Beam Search (\texttt{SBBS}), a search algorithm guided by the LM $L$ and the music emotion classifiers, denoted as $E_{m, v}$ and $E_{m, a}$, for valence and arousal. The goal of \texttt{SBBS} is to allow for the generation of pieces that sound ``good'' (i.e., have high probability value according to the trained LM), but that also match the current emotion of the story being told by the players.

% As any beam search algorithm, \texttt{SBBS} keeps in memory the set of best $b$ elements encountered in search (we formally define
% % given emotion \cite{ferreira_2019}.
% ``best'' below).
We call \texttt{SBBS} ``stochastic'' because it samples from a distribution instead of greedily selecting the best sequences of symbols, as a regular beam search does. The stochasticity of \texttt{SBBS} allows it to generate a large variety of musical pieces for the same values of $v$ and $a$.  We
also call it ``bi-objective'' because it optimizes for realism and emotion.

The pseudocode of \texttt{SBBS} is shown in Algorithm~\ref{alg:sbs}. In the pseudocode we use letters $x, y$ and $m$ to denote sequences of musical symbols. Function $p_L(y) = \prod_{y_t \in y} P(y_t|y_0, \cdots, y_{t-1})$ is the probability of sequence $y$ according to the LM $L$; a high value of $p_L(y)$ means that $y$ is recognized as a piece of
``good quality'' by $L$. We denote as $l(y)$ the duration in seconds of piece $y$. Finally, we write $x[i:j]$ for $j \geq i$ to denote the subsequence of $x$ starting at index $i$ and finishing at index $j$.

\begin{algorithm}[t]
\caption{Stochastic Bi-Objective Beam Search}
\label{alg:sbs}
\begin{algorithmic}[1]
\REQUIRE Music emotion classifier $E_m$, LM $L$, previously composed symbols $x$, valence and arousal values $v$ and $a$, number $k$ of symbols to consider, beam size $b$, length $l$ in seconds of the generated piece.
\ENSURE Sequence of symbols of $l$ seconds.
\STATE $B \gets [x]$, $j \gets 0$ \label{line:sbs:init}
% \FOR{$i = 1$ to $b$}
% \STATE $B$.append$(x \cup x_t)$ where $x_t \sim P(\cdot|x)$
% \ENDFOR
\WHILE{$l(y[t:t+j]) < l$, $\forall y \in B$} \label{line:sbs:stopping_condition}
    \STATE $C \gets \{\}$ \label{line:sbs:init_while}
    \FORALL{$m \in B$}
        \STATE $C_m \gets \{m \cup s \vert s \in V\}$ \label{line:sbs:children}
        \STATE $C_m \gets k$ elements $y$ from $C_m$ with largest $p_L(y)$ \label{line:sbs:pruning_model}
        \STATE $C \gets C \cup C_i$ \label{line:sbs:total_children}
    \ENDFOR
    \STATE $B \gets b$ sequences $y$ sampled from $C$ proportionally to $p_L(y) (1 - |v - E_{m,v}(y)|) (1 - |a - E_{m,a}(y)|)$ \label{line:sbs:sample_next_beam}
    \STATE $j \gets j + 1$ \label{line:sbs:end_while}
    % \begin{equation*}
    % p(y) \cdot (1 - |v - E_{m,v}(y)|) \cdot (1 - |a - E_{m,a}(y)|) \,.
    % \end{equation*}
\ENDWHILE
\RETURN $m \in B$ such that $p_L(m) = \max_{y \in B}p_L(y)$ and $l(y[t: t+j]) \geq l$ \label{line:sbs:return}
\end{algorithmic}
\end{algorithm}

\texttt{SBBS} initializes the beam structure $B$ with the sequence $x$ passed as input (line~\ref{line:sbs:init}).
%We store in $B$ at most $b$ sequences of symbols.
\texttt{SBBS} also initializes variable $j$ for counting the number of symbols added by the search. \texttt{SBBS} keeps in memory at most $b$ sequences and, while all sequences are shorter than the desired duration $l$ (line
~\ref{line:sbs:stopping_condition}), it adds a symbol to each sequence (lines~\ref{line:sbs:init_while}--\ref{line:sbs:end_while}). \texttt{SBBS} then generates all sequences by adding one symbol from vocabulary $V$ to each sequence $m$ from $B$ (line~\ref{line:sbs:children}); these extended sequences, known as the children of $m$, are stored in $C_m$.

The operations performed in lines~\ref{line:sbs:pruning_model} and \ref{line:sbs:sample_next_beam} attempt to ensure the generation of good pieces that convey the desired emotion. In line~\ref{line:sbs:pruning_model}, \texttt{SBBS} selects the $k$ sequences with largest $p_L$-value among the children of $m$. This is because some of the children with low $p_L$-value could be attractive from the perspective of the desired emotion and,
%, if they are sampled,
%to be part of the piece,
although the resulting piece could convey the desired emotion, the piece would be of low quality according to the LM. The best $k$ children of each sequence in the beam are added to set $C$ (line
~\ref{line:sbs:total_children}). Then, in line~\ref{line:sbs:sample_next_beam}, \texttt{SBBS} samples the sequences that will form the beam of the next iteration. Sampling occurs proportionally to the values of $p_L(y) (1 - |v - E_{m,v}(y)|) (1 - |a - E_{m,a}(y)|)$, for sequences $y$ in $C$. A sequence $y$ has higher chance of being selected if $L$ attributes a high probability value to $y$ and if the music emotion model classifies the values of valence and arousal of $y$ to be similar to the desired emotion.
When at least one of the sequences is longer than the desired duration of the piece, \texttt{SBBS} returns the sequence with largest $p_L$-value that satisfies the duration constraint (line~\ref{line:sbs:return}).

\section{Empirical Evaluation}

Our empirical evaluation is divided into two parts. First, we evaluate the accuracy of the models used for story and music emotion classification. We are interested in comparing the fine-tuned BERT model for story emotion classification with the simpler Na\"ive Bayes approach of \citeauthor{padovani2017}~\shortcite{padovani2017}. We are also interested in comparing the fine-tuned GPT-2 model for music emotion classification with the simpler LSTM of \citeauthor{ferreira_2019}~\shortcite{ferreira_2019}. In the second part of our experiments we evaluate with a user study whether human subjects can recognize different emotions in pieces generated by Composer for the CotW campaign.

% We first evaluate the classification
% accuracies of the story's emotion classifier as well as the
% music emotion classifier.

\subsection{Emotion Classifiers}

\subsubsection{Story Emotion}

The story emotion classifier we use with Composer is a pair of BERT models, one for valence and one for arousal. For both models, we use the pre-trained BERT$_{BASE}$ of \citeauthor{devlin2018bert}~\shortcite{devlin2018bert},
which has 12 layers, 768 units per layer, and 12 attention heads. BERT$_{BASE}$ was pre-trained using both the BooksCorpus (800M words) \cite{zhu2015aligning} and
the English Wikipedia (2,500M words).

We independently fine-tune these two BERT models as valence and arousal
classifiers using the CotW dataset \cite{padovani2017}. Fine-tuning consists of adding a
classification head on top the pre-trained model and training all the
parameters (including the pre-trained ones) of the resulting model
end-to-end. All these parameters were fine-tuned with an Adam optimizer~\cite{adam14} with learning rate of 3e-5 for 10 epochs. We used mini-batches of size 32 and dropout of 0.5.

The CotW dataset is divided into 9 episodes, thus we evaluate
the accuracy of each BERT classifier using a leave-one-out strategy.
For each episode $e$, we leave $e$ out for
testing and train in the remaining episodes. For example,
when testing on episode 1, we use episodes 2-8 for training.
Every sentence is encoded using a
WordPiece embedding \cite{wu2016google} with a 30,000 token vocabulary.

We compare the fine-tuned BERT classifiers with a Na\"ive
Bayes (NB) approach (baseline), chosen because it is the method underlying the original Bardo system. NB encodes sequences using a traditional bag-of-words with tf–idf approach.
Table~\ref{tab:valence} shows the accuracy of the valence classification of both these methods per episode. The best accuracy for a given episode is highlighted in bold. The BERT classifier outperforms NB in all the episodes, having an average accuracy 7\% higher.
For valence classification, the hardest episode for both the models is episode 7, where BERT had the best performance improvement when compared to NB. The story told in episode 7 of CotW is different from all other episodes. While the other episodes are full of battles and ability checks, episode 7 is mostly the players talking with non-player characters.
%while convincing them to join their party in an upcoming war.
Therefore, what is learned in the other episodes does not generalize well to episode 7. The improvement in accuracy of the BERT model in that episode is likely due to the model's pre-training. Episodes 5 and 9 were equally easy for both methods because they are similar to one another. The system trained in one of these two episodes generalizes well to the other.

\begin{table}[!t]
\centering
%\tiny
%\small
\setlength{\tabcolsep}{4pt}
\begin{tabular}{crrrrrrrrrr}
%\cline{2-11}
\toprule
\multirow{2}{*}{\textbf{Alg.}} & \multicolumn{9}{c}{\textbf{Episodes}} & \multirow{2}{*}{\textbf{Avg.}} \\
\cmidrule{2-10}
& \multicolumn{1}{c}{\textbf{1}}   & \multicolumn{1}{c}{\textbf{2}}   & \multicolumn{1}{c}{\textbf{3}}  & \multicolumn{1}{c}{\textbf{4}} & \multicolumn{1}{c}{\textbf{5}}  & \multicolumn{1}{c}{\textbf{6}}  & \multicolumn{1}{c}{\textbf{7}} & \multicolumn{1}{c}{\textbf{8}}  & \multicolumn{1}{c}{\textbf{9}} &    \\
\midrule
\multicolumn{1}{l}{\textbf{NB}}   &   73 & 88  &  91 & 85   &  94 & 81  &  41 & 74   & 94    &   80 \\
\multicolumn{1}{l}{\textbf{BERT}}   &  \textbf{89}  & \textbf{92}  & \textbf{96}  &  \textbf{88} & \textbf{97}   & \textbf{81} & \textbf{66}   &  \textbf{83}   &  \textbf{96} &  \textbf{87}  \\
\bottomrule
\end{tabular}
\caption{Valence accuracy in \% of Na\"ive Bayes (NB) and BERT for story emotion classification.}
\label{tab:valence}
\end{table}

Table \ref{tab:arousal} shows the accuracy of arousal classification
of both NB and BERT. The best accuracy for a given episode is highlighted in bold. Again BERT outperforms NB in all the episodes, having an average accuracy 5\% higher. In contrast with the valence results, here there is no episode in which the BERT model substantially outperforms NB.

\begin{table}[!h]
\centering
%\tiny
%\small
\setlength{\tabcolsep}{4pt}
\begin{tabular}{crrrrrrrrrr}
%\cline{2-11}
\toprule
\multirow{2}{*}{\textbf{Alg.}} & \multicolumn{9}{c}{\textbf{Episodes}} & \multirow{2}{*}{\textbf{Avg.}} \\
\cmidrule{2-10}
& \multicolumn{1}{c}{\textbf{1}}   & \multicolumn{1}{c}{\textbf{2}}   & \multicolumn{1}{c}{\textbf{3}}  & \multicolumn{1}{c}{\textbf{4}} & \multicolumn{1}{c}{\textbf{5}}  & \multicolumn{1}{c}{\textbf{6}}  & \multicolumn{1}{c}{\textbf{7}} & \multicolumn{1}{c}{\textbf{8}}  & \multicolumn{1}{c}{\textbf{9}} &    \\
\midrule
\multicolumn{1}{l}{\textbf{NB}}   &   82 & 88  &  75 & 79   &  82 & 76  &  98 & 86   & 84    &   83 \\
\multicolumn{1}{l}{\textbf{BERT}}   &  \textbf{86}  & \textbf{90}  & \textbf{77}  &  \textbf{86} & \textbf{89}   & \textbf{88} & \textbf{99}   &  \textbf{90}   &  \textbf{88} &  \textbf{88}  \\
\bottomrule
\end{tabular}
\caption{Arousal accuracy in \% of Na\"ive Bayes (NB) and BERT for story emotion classification.}
\label{tab:arousal}
\end{table}

\subsubsection{Music Emotion}

The music emotion classifier is a pair of GPT-2 models, one for valence and one for arousal.
We first pre-trained a GPT-2 LM with our ADL Piano MIDI dataset. We augmented each piece $p$ of this dataset  by (a) transposing $p$ to every key, (b)
increasing and decreasing $p$'s tempo by 10\% and (c) increasing and decreasing
the velocity of all notes in $p$ by 10\% \cite{oore2017learning}. Thus, each
piece generated $12 \cdot 3 \cdot 3 = 108$ different examples.

The pre-trained GPT-2 LM has 4 layers (transformer blocks), context size of 1024
tokens, 512 embedding units, 1024 hidden units, and 8 attention heads.  We then fine-tuned the GPT-2 LM independently using the VGMIDI dataset, for valence
and for arousal. Similarly to BERT, fine-tuning a GPT-2
architecture consists of adding an extra classification head on top of the pre-trained model and training all parameters end-to-end. Similar to the story emotion classifiers, we
fine-tuned the GPT-2 classifiers for 10 epochs using an Adam optimizer with learning rate 3e-5. Different from the story emotion classifiers, we used mini-batches of size 16 (due to GPU memory constrains) and dropout of 0.25. The VGMIDI dataset is defined with a train and test splits of 160 and 40 pieces, respectively. We augmented the dataset by slicing each piece
into 2, 4, 8 and 16 parts of equal length and emotion. Thus, each part of each slicing
generated one extra example. This augmentation is intended
to help the classifier generalize for pieces
with different lengths.
%We used the same splits to evaluate the GPT-2 classifiers for valence and arousal.

We compare the fine-tuned GPT-2 classifiers with LSTM models that were also
pre-trained with the ADL Piano Midi dataset and fine-tuned with the VGMIDI dataset. We chose LSTMs because they are the state-of-the-art model in the VGMIDI dataset~\cite{ferreira_2019}. The LSTMs have same size as the GPT-2 models (4 hidden layers, 512 embedding units, 1024 hidden units) and were pre-trained and fine-tuned with the same hyper-parameters.
Table \ref{tab:sent_accuracy} shows the accuracy of both models for valence and arousal. We also report the performance of these models
without pre-training (i.e., trained only on the VGMIDI dataset). We call
these the baseline versions of the models.

\begin{table}[!t]
    \centering
    \begin{tabular}{ccc}
    \toprule
    \textbf{Algorithm} & \textbf{Valence} & \textbf{Arousal} \\
    \midrule
    Baseline LSTM & 69 & 67 \\
    Fine-tuned LSTM & 74 & 79 \\
    Baseline GPT-2 & 70 & 76 \\
    Fine-tuned GPT-2 & \textbf{80} & \textbf{82} \\
    \bottomrule
    \end{tabular}
    \caption{Accuracy in \% of both the GPT-2 and LSTM models for music emotion classification. }
    \label{tab:sent_accuracy}
\end{table}

Results show that using transfer learning can substantially boost the performance
of both models. The fine-tuned GPT-2 is 10\% more accurate in terms of valence and 8\% in terms of arousal. The fine-tuned LSTM is 5\% more accurate in terms of
valence and 12\% in terms of arousal. Finally, the fine-tuned GPT-2
outperformed the fine-tuned LSTM by 6\% and 3\% in terms of valence and arousal, respectively. % with respect to valence and by 3\% with respect to arousal.

\begin{table*}[!t]
\centering
%\tiny
%\small
\setlength{\tabcolsep}{4pt}
\begin{tabular}{crrrrrrrrrrrrrrrrrrrrrrr}
%\cline{2-11}
\toprule
\multirow{3}{*}{\textbf{Method}} & \multicolumn{20}{c}{\textbf{Episodes}} & \multicolumn{3}{c}{\multirow{3}{*}{\textbf{Average}}} \\
\cmidrule{2-21}
% & \multicolumn{4}{c}{\textbf{1}} & \multicolumn{4}{c}{\textbf{2}}   & \multicolumn{4}{c}{\textbf{3}} & \multicolumn{4}{c}{\textbf{4}} & \multicolumn{4}{c}{\textbf{5}} &    \\
& \multicolumn{2}{c}{\textbf{e1-p1}} & \multicolumn{2}{c}{\textbf{e1-p2}} & \multicolumn{2}{c}{\textbf{e2-p1}} & \multicolumn{2}{c}{\textbf{e2-p2}} & \multicolumn{2}{c}{\textbf{e3-p1}} & \multicolumn{2}{c}{\textbf{e3-p2}} & \multicolumn{2}{c}{\textbf{e4-p1}} & \multicolumn{2}{c}{\textbf{e4-p2}} & \multicolumn{2}{c}{\textbf{e5-p1}} & \multicolumn{2}{c}{\textbf{e5-p2}} \\
& \textbf{v} & \textbf{a} & \textbf{v} & \textbf{a} & \textbf{v} & \textbf{a} & \textbf{v} & \textbf{a} & \textbf{v} & \textbf{a} & \textbf{v} & \textbf{a} & \textbf{v} & \textbf{a} & \textbf{v} & \textbf{a} & \textbf{v} & \textbf{a} & \textbf{v} & \textbf{a} & \textbf{v} & \textbf{a} & \textbf{va}\\
\cmidrule{2-21}
\multicolumn{1}{l}{\textbf{Baseline}} & 56 & \textbf{65} & 39 & 56 & 39 & 62 & 39 & \textbf{79} & \textbf{48} & \textbf{60} & \textbf{67} & \textbf{53} & \textbf{58} & 70 & \textbf{63} & \textbf{75} & 25 & 36 & \textbf{72} & 58 & \textbf{51} & \textbf{32} & \textbf{34}\\
\multicolumn{1}{l}{\textbf{Composer}} & \textbf{62} & 60 & \textbf{44} & \textbf{65} & \textbf{82} & \textbf{68} & \textbf{53} & 68 & 24 & 55 & 46 & 43 & 25 & \textbf{87} & 37 & 55 & \textbf{81} & \textbf{86} & 51 & \textbf{67} & \textbf{51} & 30 & \textbf{34}\\
\bottomrule
\end{tabular}
\caption{The percentage of participants  that  correctly  identified  the valence and arousal (v and a, respectively) intended by the methods for the pieces parts (p1 and p2).
%The table also reports the average accuracy for all generated pieces in terms of valence (v), arousal (a), and jointly for valence and arousal (va).
}
\label{tab:user_study}
\end{table*}

\subsection{User Study}

% We evaluate Composer with a user study where the participants
% listened to pieces generated by the system.
In our study we measure Composer's performance at generating music
that matches the emotions of a story. We use Composer to generate
a piece for a snippet composed of 8 contiguous sentences of each of the first 5 episodes of the CotW dataset.
%Each snippet is composed of 8 contiguous sentences.
Each snippet has one emotion transition that happens in between sentences. The sentences are 5.18 seconds long on average.
% We ensured there was a transition to be able to evaluate more emotions in a single excerpt.
% As defined in Algorithm
% \ref{alg:bardo}, Composer
% %processes these 8 sentence one-by-one,
% %generating
% generates
% one music piece for each sentence.
% To generate the piece for a sentence $t$, Composer considers
% the emotion of $t$ (as given by the story emotion classifier) and
% the piece generated for sentence $t-1$. Thus, when generating a piece
% for sentence $t$, we expect Composer to continue the piece
% from sentence $t-1$, accounting for potential changes in valence and
% arousal.
To test Composer's ability to generate
%these smooth continuations
music pieces with emotion changes, we asked human subjects to listen
to the 5 generated pieces and evaluate the transitions of emotion in
each generated piece.\footnote{Generated pieces can be downloaded from the following link: \url{https://github.com/lucasnfe/bardo-composer}}

The user study was performed via Amazon Mechanical Turk and had an expected completion time of approximately 10 minutes. A reward
of USD \$1 was given to each participant who completed the
study.
% In the first section of the study, subjects were
% presented with information about the experiment such as the authors
% affiliation, goals of the experiment, tasks to be performed,
% duration of the tasks, and data collection privacy.
In the
first section of the study, the participants were presented an illustrated description
of the valence-arousal model of emotion and listened to 4 examples of pieces
from the VGMIDI dataset labeled with the valence-arousal model. Each piece had a different emotion: low valence and arousal, low valence and high arousal, high valence and low arousal, high valence and arousal.

In the second section of the study, participants were asked to
listen to the 5 generated pieces (one per episode). After listening to each piece, participants had
to answer 2 questions: (a) ``What emotion do you perceive in the 1st part of the piece?'' and (b) ``What emotion do you perceive in the 2nd part of the piece?'' To answer these two questions, participants selected one of the
four emotions: low valence and arousal, low valence and high arousal, high valence and low arousal, high valence and arousal.
% To answer
% the third question, participants used a Likert scale from 1 to 5 where
% 1 means ``Not smooth at all'' and 5 means ``very smooth''.
Subjects were allowed to play the pieces as many times as they
wanted before answering the questions.
The final section of the study was a demographics questionnaire including ethnicity, first language, age, gender, and
experience as a musician. To answer the experience as a musician,
we used a 1-to-5 Likert scale where 1
means ``I've never studied music theory or practice'' and 5 means ``I
have an undergraduate degree in music''.

We compare Composer with a baseline method that selects a random piece from the
VGMIDI dataset whenever there is a transition of emotion. The selected piece has
the same emotion of the sentence (as given by the story emotion classifier). To compare these two methods, we used a
between-subject strategy where Group $A$ of 58 participants evaluated the
5 pieces generated by Composer and another Group $B$ of 58 participants
evaluated the 5 pieces from the baseline.
We used this
strategy to avoid possible learning effects where subjects
could learn emotion transitions from one method and apply the same evaluation directly to the other method. The average age of
groups $A$ and $B$ are 34.96 and 36.98 years, respectively. In Group
$A$, 69.5\% of participants are male and 30.5\% are female. In Group
$B$, 67.2\% are male and 32.8\% are female. The average musicianship of the groups $A$ and $B$ are 2.77 and 2.46, respectively.

Table \ref{tab:user_study} shows the results of the user study.
% Considering the accuracy on the combined dimensions (VA) in episodes 2 and 5 (i.e. the system matched both the valence and arousal for that sentence), Composer outperformed the Baseline by approximately 50\% .  In episodes 3 and 4, the Baseline outperformed the Composer by approximately the same percentage. In episode 1, both systems performed
% similarly, with the baseline having a slightly advantage.
We consider both parts (p1 and p2 in the table) of each episode as an independent piece.
%in
%order to analyse the results more clearly.
% The table presents the percentage of participants that correctly identified an approach's intended emotion for its musical piece,
The table presents the percentage of participants that correctly identified the pieces' valence and arousal (``v'' and ``a'' in the table, respectively), as intended by the methods.
For example, 87\% of the participants correctly identified the arousal value that Composer intended the generated piece for part p1 of episode 4 (e4-p1) to have. We refer to the percentage of participants that are able to identify the approach's intended emotion as the approach's accuracy.
%In the ``Average'' section of the table,
We also present the approaches' average accuracy across all pieces (``Average'' in the table) % generated by Composer and by Baseline
%(professional musicians)
in terms of valence, arousal, and jointly for valence and arousal (``va'' in the table). The ``va''-value of 34 for Composer means that 34\% of the participants correctly identified the system's intended values for valence and arousal across all pieces generated. % by the system.

Composer outperformed the Baseline in e1-p2, e2-p1, and e5-p1. Baseline outperformed
Composer e3-p1, e3-p2 and e4-p2. In the other four parts,
one method performed better for valence whereas
the other method performance better for arousal.
Overall, the average results show that both systems performed very
similarly. Both of them had an average accuracy on the combined
dimensions equal to 34\%. The difference between these two methods and a
system that selects pieces at random (expected accuracy of 25\%)
is significant according to a Binomial test ($p = 0.02$).
%This shows that both systems performed approximately
%10\% better than a random system.
These results show that the  participants  were  able  to  identify  the  emotions  in  the generated pieces as accurately as they were able to identify the emotions in human-composed pieces. This is an important
result towards the development of a fully automated system for music composition for story-based tabletop games.

\section{Conclusions}

This paper presented Bardo Composer, a system that
automatically composes music for tabletop role-playing games. The system processes sequences
from speech and generates pieces one sentence after the other. The emotion of the sentence is classified using a fine-tuned BERT. This emotion is given as
input to a Stochastic Bi-Objective Beam Search algorithm that tries to generate a piece that matches the emotion.
% We evaluate the accuracy of the BERT classifier and results showed that it outperformed a Na\"ive Bayes
% approach.
We evaluated Composer with a user study and results showed that human subjects correctly identified the emotion of the generated music pieces as accurately as they were able to identify the emotion of pieces composed by humans.
% In the future we intend to evaluate Bardo Composer
% as a system for generating background music for
% live tabletop RPG game sessions.

\bibliography{bardo}

\end{document}